\documentclass[conference]{IEEEtran}
\IEEEoverridecommandlockouts
\usepackage{cite}
\usepackage{amsmath,amssymb,amsfonts}
\usepackage{algorithmic}
\usepackage{graphicx}
\usepackage{textcomp}
\usepackage{xcolor}
\usepackage[utf8]{inputenc}
\usepackage{float}
\def\BibTeX{{\rm B\kern-.05em{\sc i\kern-.025em b}\kern-.08em
    T\kern-.1667em\lower.7ex\hbox{E}\kern-.125emX}}
\usepackage{array,multirow}
\usepackage{booktabs}
\usepackage{pifont}
\newcommand*\rot{\rotatebox{90}}

\title{An Overview of Federated Deep Learning Privacy Attacks and Defensive Strategies
}

\usepackage{balance}

\author{
\IEEEauthorblockN{David Enthoven \ \ \ \ \ \ \ \ Zaid Al-Ars}
\IEEEauthorblockA{\textit{Accelerated Big Data Systems Group} \\
\textit{Delft University of Technology, NL}\\
\{D.A.Enthoven, Z.Al-Ars\}@tudelft.nl}
}

\begin{document}
\maketitle

\begin{abstract}
With the increased attention and legislation for data-privacy, collaborative machine learning (ML) algorithms are being developed to ensure the protection of private data used for processing. Federated learning (FL) is the most popular of these methods, which provides privacy preservation by facilitating collaborative training of a shared model without the need to exchange any private data with a centralized server. Rather, an abstraction of the data in the form of a machine learning model update is sent. Recent studies showed that such model updates may still very well leak private information and thus a more structured risk assessment is needed. In this paper, we analyze existing vulnerabilities of FL and subsequently perform a literature review of the possible attack methods targeting FL privacy protection capabilities. These attack methods are then categorized by a basic taxonomy.  Additionally, we provide a literature study of the most recent defensive strategies and algorithms for FL aimed to overcome these attacks. These defensive strategies are categorized by their respective underlying defense principle. The paper concludes that the application of a single defensive strategy is not enough to provide adequate protection to all available attack methods.\\
\end{abstract}


\section{Introduction} \label{sec:intro}

Deep learning algorithms have grown significantly in capabilities and popularity in the past decade. Advances in computing performance, increased data availability (and decreased data-storage costs) and advances in more effective algorithms have created a surge of new applications using deep learning algorithms.

For these algorithms to be accurate, generally a large amount of task-specific data is needed. Conventionally, this data is gathered and stored on an accessible centralized server, which is becoming increasingly cheap to access due to the increasing bandwidths available for remote data storage and retrieval.

This centralized data storage, however, introduces two problems. First, due to the exponential increase of data collectors in the field, the amount of data that is collected is becoming excessively large, which in turn increases the bandwidth requirements for communication as well as the computational requirements needed for processing. Second, devices may gather data of sensitive nature which is subject to privacy laws and regulations~\cite{GDPR}. Therefore, it is important to manage this sensitive data in a way that protects privacy. To combat these problems alternative ways of decentralized learning have been proposed, most notably that of federated learning.

\begin{figure}[t]
    \centering
    \includegraphics[width=\linewidth]{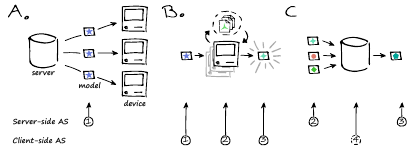}
    \caption{\emph{Top}: The Federated learning process: A) A model is sent from the servers to the devices. B) The devices train the model further on their local dataset. C) The devices send their model back to the server which combines them into a singular model.\newline
    \emph{Bottom}: The points of vulnerability defining the attack surface (AS) for server-side and client-side attack methods.}
    \label{fig:fed_steps}
\end{figure}

FL~\cite{FedLearning} is a specific form of distributed learning methods~\cite{SparkNet} which relies on the notion that no actual data need to be sent to a centralized server. Rather, an abstraction in the form of machine learning models is sent to the server instead. These models are trained locally on client-side devices with the clients' datasets. These updated models are sent back to the centralized server. Then, the server aggregates these models into a single model and sends this model back to the clients. This process is illustrated in Figure~\ref{fig:fed_steps}. As this process happens multiple times iteratively, all clients jointly train the shared model. FL claims to have distinct privacy advantages over traditional centralized data-driven model training approaches as well as reducing communication bandwidth requirements~\cite{FedLearning}. These privacy advantages are focused on protecting sensitive data from external adversaries, but it does not guarantee safety against internal adversaries such as malicious servers. 

FL has been demonstrated to outperform traditional training methods in certain cases. Researchers demonstrated an implementation for next word prediction on mobile devices~\cite{Fed_nextword,Fed_nextword2} as well as emoticon prediction~\cite{Fed_emoticon} among others. Furthermore, FL has found its use in medical applications as well~\cite{eurocat,medfed1,medfed2,medfed3,medfed4,medfed5} which, due to the sensitive nature of medical data, creates a need for justification of the safety of FL. This safety comes two-fold: First, FL should protect against adversarial clients. Second, due to modern information storage legislation (such as the General Data Protection Regulation or GDPR~\cite{GDPR}) private user data may not be gathered without the users' consent. Therefore FL must be shown to be secure against the server obtaining private user information from the user models.

However, recent research has shown multiple vulnerabilities in the privacy protection capabilities of FL~\cite{threatsurvey1,DLG,iDLG,First-layer-attack,warfarin,MIA1,mGAN-AI,GAN-attack,backdoor2,Backdoorattack,backdoor3,sybilattack}. These attack methods cast a shadow on the applicability of FL for privacy-sensitive information.

In this paper, we present an overview of attack methods known to affect FL and provide a basic taxonomy to classify these methods. Furthermore, we provide a literature review of defensive strategies against such attack methods which we categorize by their respective underlying defensive strategy. This will help researchers to appropriately weigh the capabilities and risks involved in using FL for privacy-sensitive information.

This paper is organized as follows. Section~\ref{sec:threat} introduces the concept of adversarial attacks in the FL context. Section~\ref{sec:attack} presents the FL attack methods already published in the literature. Then Section~\ref{sec:defense} discusses defensive strategies able to counteract possible attack methods. Section~\ref{sec:related} outlines related work already published on the topic. Finally, Section~\ref{sec:conc} ends with the conclusions.

\begin{figure*}
    \centering
    \includegraphics[width=0.7\linewidth]{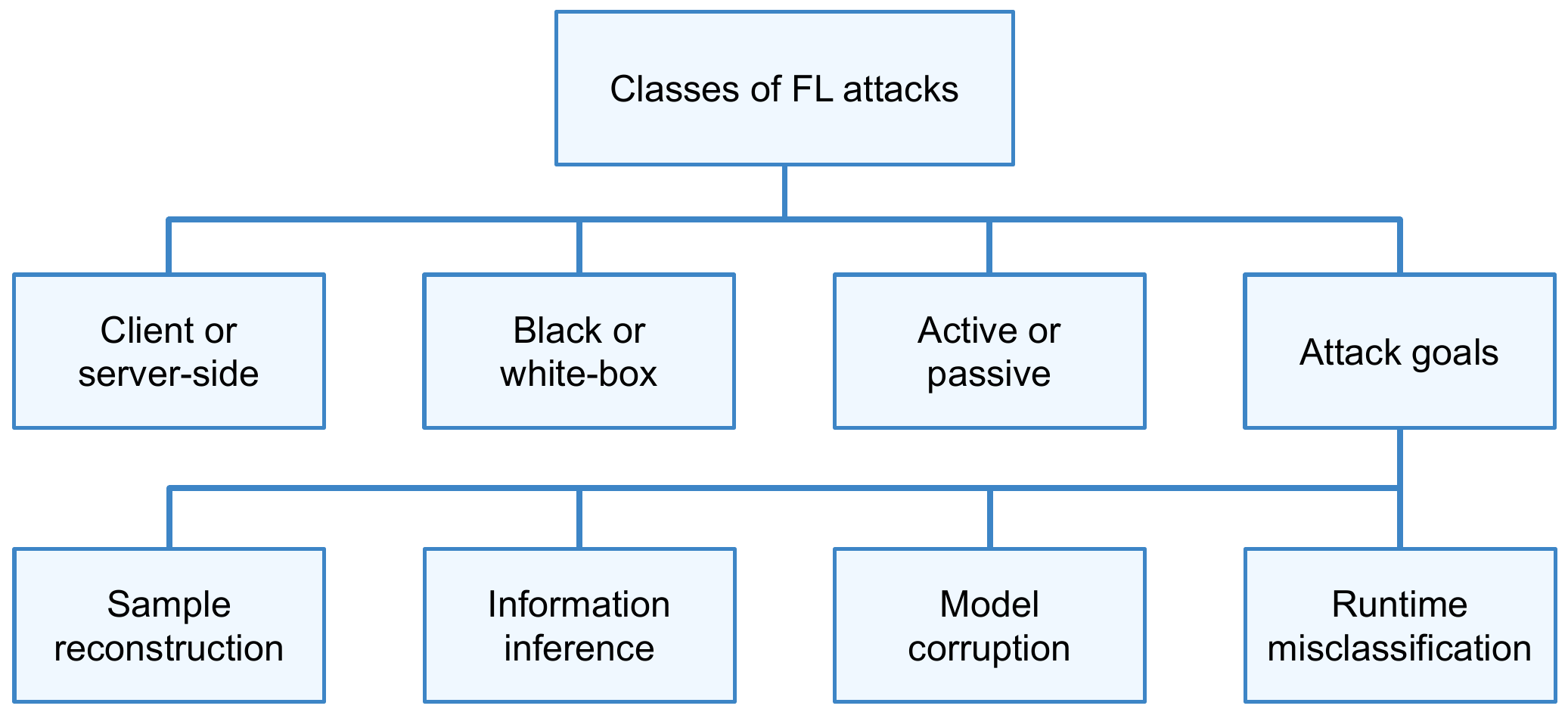}
    \caption{Taxonomy to classify the different types of FL attack methods}
    \label{fig:classes}
\end{figure*}

\section{Threat model} \label{sec:threat}

\subsection{Attack surface}
In general, an adversary attacking an FL model has one of two adversarial goals. First and foremost is to extract private information from victim clients. A second goal is to purposefully force the model to behave differently than intended. This can take the form of introducing backdoors, inducing miss-classification or even rendering the model unusable.

Throughout this paper, an adversary is considered to be on the inside, meaning it could be one of the clients participating in the FL system or could be the centralized server itself. This consideration is justified by the assumption that all communication between the server and the different clients can be done with the latest cryptographic security. One of the key principals for which FL was developed is that the server is not allowed to have access to any of the private training samples of the clients. Therefore, the honest-but-curious server model is considered in the threat model as well. Unless stated otherwise, the basic form of FL is the one used in the assessment of the threats.

The attack surface is the summation of exploitable vulnerabilities in the system. The viable attack surface in FL is subdivided for the case of a malicious client and malicious server. The vulnerable points of attack of FL are illustrated in the bottom part of Figure~\ref{fig:fed_steps}. 

The attack surface for a malicious server consists of the following exploitable points. 1) The ability to adapt the model sent to all clients and/or ability to determine and regulate the participating clients. Furthermore, the server can send custom models to targeted clients. 2) The ability to differentiate the model updates from clients before aggregation. 3) The server is knowledgeable of the aggregate model.

The attack surface for malicious clients consists of the following usable exploits. 1) The client has access to the aggregate model. 2) The ability to manipulate the data on which the client trains the model, as well as regulating the training process. 3) The ability to manipulate their gradient update. 4) The ability to influence the impact of their model during aggregation.

There are several ways to characterize FL attacks based on their source, nature, etc. Figure~\ref{fig:classes} shows the taxonomy of the different characteristics we use in this paper to describe FL attacks. In the following, each one the shown taxonomy classes is discussed, along with the general vulnerabilities FL has.

\subsection{Client or server-side attacks} 
In this paper, all communication between the server and the clients is assumed to be secure. Only the clients and server themselves are considered viable points for the execution of attack methods. Therefore, the system is only considered to be vulnerable either on the client-side or the server-side. The one exception to this rule is the man-in-the-middle attack, where an adversary pretends to be a server to the client or/and a client to the server and is therefore in-between the communication. Such an adversary is assumed to have the same threat model as for server-side vulnerabilities.

\subsection{Black-box or white-box attacks} 
White-box attacks mean that the adversary has complete knowledge of the system except for the private data. Black-box attacks mean that the adversary is limited and is only able to use the system without any detailed knowledge of the inner workings of the system.

In the federated averaging algorithm, the clients are assumed to have a full description of the model and are therefore vulnerable to white-box attacks. On the server-side, this is also true but recent studies propose models in which the server has no knowledge of the model other than its conception~\cite{Cronus}. This might force the server-side attacks towards a black-box scenario.

In some cases, it is possible to extract information about the systems architecture and thereby force the black-box towards a more white-box threat model. Method such as GPU-snooping can be exploited~\cite{ModelExtraction} or model extraction for online services~\cite{StealingModel}.

\subsection{Active or passive attacks} 
\emph{Active attacks} imply that the attack method relies on manipulating the system by adapting parameters, communication or other system-properties to achieve adversarial goals. Because these types of attack influence the working system they are in a general sense detectable. Active attacks may also be responsible for corruption the system to a point where it is no longer properly functioning.

\emph{Passive attacks} imply that the attack method can be performed without the need to adapt any parameters in the system. Since these attack methods leave no trace of their execution, they are generally impossible to detect, which makes them a more dangerous set of attack methods.

\subsection{Attacker goal}
Each attack should be classified by the goal of the adversary. The attacks presented in this paper fall in one of the following four adversarial goals.

\emph{1.~Sample reconstruction} aims to objectively reveal what training data was used by the clients. The success of the attacker in achieving this goal can be assessed by comparing the similarity of the reconstruction with the original data. The ability to launch a reliable sample reconstruction attack is the most intrusive concerning private information since it reveals sensitive information in full. If a sample reconstruction attack can be carried out with relative ease by the server or an adversary, the system as a whole may not claim to be privacy-preserving.

\emph{2.~Information inference} attacks aim to learn private information more speculatively. This information may consist of:
\begin{itemize}
\item \emph{Inferring class representatives}---This type of inference tries to recreate samples that would not stand out in the original dataset. When successfully creating such samples the adversary can learn a lot about the underlying dataset.
\item \emph{Inferring membership (of the sample)}---When provided a sample, membership inference tries to accurately determine if this sample has been used for the training of the network.
\item \emph{Inferring data properties}---This attack tries to infer meta-characteristics of the used dataset (e.g.~the data consists of mostly red cars).
\item \emph{Inferring samples/labels}---This attack aims to accurately recreate (rather than reconstruct) training-samples that were used during the training session and/or associated labels that were used during training.
\end{itemize}

\emph{3.~Model corruption} aims to change the model to the benefit of the attacker. Such attacks can introduce exploitable backdoors in the model or even make the entire model useless. Model corruption attacks are generally not assumed to violate the privacy of participating clients.

\emph{4.~Runtime misclassification} attacks aim to trick the model into making erroneous predictions during runtime. FL is susceptible to these types of attacks similar to non-distributed ML methods. Generally, runtime misclassification attacks do not violate client privacy directly but may be used to circumvent ML-based security measures.

\begin{table*}[!ht]
\small
\caption{FL attack methods and their characteristics}
\label{tab:attackmethods}
\begin{center}\begin{tabular}{l | l l l l l}
Attack method               &Attacker   
&\begin{tabular}{@{}l@{}}Active/\\passive\end{tabular}         
&\begin{tabular}{@{}l@{}}White-box /\\Black-box \end{tabular} 
&Attack goal                &Restricted to\\

\hline
Loss-function/ReLu          &Server            &Passive         &White-box          &Sample reconstruction      &Linear net + ReLu \\
First dense layer           &Server            &Passive         &White-box          &Sample reconstruction      &Non-recurrent, small local dataset\\
DLG/iDLG                    &Server            &Passive         &White-box          &Sample reconstruction      &Small local dataset\\
MIA                         &Server            &Passive         &Both               &Inference                  &Linear nets, small input space\\
mGAN-AI                     &Server            &Both            &White-box          &Inference                  & \begin{tabular}[t]{@{}l@{}}Requires auxiliary dataset, synthesizable data \end{tabular}\\
GAN                         &Client            &Active          &White-box          &Inference                  &Sparse client-set\\
Adversarial example         &Client            &Passive         &White-box          &Misclassification          &Limited practical use\\
Model poisoning             &Client            &Active          &Both               &Model corruption, backdoor &
\end{tabular}
\end{center}\end{table*}

\section{Attack methods} \label{sec:attack}

Table~\ref{tab:attackmethods} lists the attack methods proposed in the literature as well as their characteristics. In the following, we discuss each one of these methods, classified according to their associated attack goals. Table~\ref{tab:ASmatrix_attack} illustrates the exploited vulnerabilities of the hereafter listed attack methods.

\subsection{Attacks targeting reconstruction}

\subsubsection{Loss-function/ReLu exploitation}
Sannai~\cite{mathattack} describes a mathematical framework that (given access to a white-box model) can determine what relation the input has to a given loss at the output. It relies on the non-smoothness of the ReLu function and has been argued to work solely on deep neural networks. Due to the use of ReLu activation on each layer, an adversary can compute backwards which nodes were activated. Several polynomials that describe the input-output relation can be formulated and subsequently, these polynomials are used to calculate what training samples are consistent with the loss surface at the output. 

The application domain of this attack method is limited to linear models with incorporated ReLu activation functions. Furthermore, this method is weak against any form of noise due to its strict mathematical nature. Unfortunately, no demonstration is given of the workings of this attack method.

\subsubsection{First dense layer attack}
Le Trieu et al.~\cite{First-layer-attack} showed mathematically that gradients of a model may in some cases reveal the training data. This attack relies on the fact that the gradients of all weights accumulated in a neuron are scaled linearly to the activation of the prior layer~\cite{homomorphic, First-layer-attack}. In the case that a client has trained on a singular sample, this method fully reconstructs this trained sample.

This attack method is not applicable to recurrent neural networks and requires extra algorithms to be useful in for convolutional networks. Furthermore, as the local dataset grows larger the reconstructions that are to be performed with this method will become more unreliable. Still, this remains a very viable attack method due to its ease of use and near-constant computational cost. 

\subsubsection{DLG/iDLG}
A recent algorithm called Deep Leakage from Gradients (DLG) describes an optimization algorithm that reproduces the training sample iteratively by optimizing the input to produce the correct gradients for a client. A slight variation to this algorithm also proves to be viable for slightly larger batches of samples~\cite{DLG}. In a follow-up paper, a different research group proposes improved DLG (iDLG) as a more efficient and accurate version of DLG~\cite{iDLG}. DLG and iDLG are applicable only when the local dataset of a client is sufficiently small due to the required high computation costs for reliable results on larger batches.

\begin{table*}[ht]
\caption{Exploited vulnerabilities per attack. The vulnerabilities are illustrated in Figure~\ref{fig:fed_steps}}
\label{tab:ASmatrix_attack}
\centering
\begin{tabular}{@{} cl*{3}c|*{4}c | @{}}
&& \multicolumn{3}{c}{Server}&\multicolumn{4}{c}{Client} \\
&\multicolumn{1}{l|}{}&
\rot{\begin{tabular}{@{}l@{}}1) Manipulate\\ model per client\end{tabular}}&
\rot{\begin{tabular}{@{}l@{}}2) Access to client-\\specific model\end{tabular}}&
\rot{\begin{tabular}{@{}l@{}}3) Access to\\ aggregate model\end{tabular}}&
\rot{\begin{tabular}{@{}l@{}}1) Access to\\ aggregate model\end{tabular}}&
\rot{\begin{tabular}{@{}l@{}}2) Train on\\ adversarial data\end{tabular}}&
\rot{\begin{tabular}{@{}l@{}}3) Manipulate\\ gradient update\end{tabular}}&
\rot{\begin{tabular}{@{}l@{}}4) Influence\\ aggregation\end{tabular}}\\ \cline{2-9} 

\begin{tabular}{@{}l@{}}\end{tabular} 
\multirow{8}{*}{\rot{Attacks}}
&\multicolumn{1}{l|}{Loss-function/Relu}        &           &\ding{51}  &\ding{51}  &           &           &           &\\
&\multicolumn{1}{l|}{First dense layer}         &           &\ding{51}  &\ding{51}  &           &           &           &\\
&\multicolumn{1}{l|}{DLG/iDLG}                  &           &\ding{51}  &\ding{51}  &           &           &           &\\
&\multicolumn{1}{l|}{MIA}                       &           &\ding{51}  &\ding{51}  &           &           &           &\\
&\multicolumn{1}{l|}{mGAN-AI}                   &\ding{51}  &\ding{51}  &\ding{51}  &           &           &           &\\
&\multicolumn{1}{l|}{GAN}                       &           &           &           &           &           &\ding{51}  &\\ 
&\multicolumn{1}{l|}{Adversarial example}       &           &           &           &\ding{51}  &           &           &\\
&\multicolumn{1}{l|}{Model poisoing}            &           &           &           &           &\ding{51}  &\ding{51}  &\ding{51}\\ 
\cline{2-9}
\end{tabular}
\end{table*}



\subsection{Attacks targeting inference}

\subsubsection{Model inversion attacks (MIA)}
The model inversion attack introduced by Fredrikson et al.~\cite{warfarin} works on the premise that the trained linear model is accessible to the adversary in a black-box fashion. The adversary may only provide its input and collect the respective output. Also, the adversary has incomplete information about the input and complete information about the output. This data is used to detect correlations between the (still unknown) input variables and the known output values~\cite{warfarin}. It does this rather crudely by 'brute forcing' all variations of unknown input-values to predict the most likely private feature. A practical application is knowing a victim's personalized outcome of a model, as well as some publicly known features such as age, gender, etc. These features allow the adversary to reasonably predict the missing data. 

A follow-up work presents a new version that has a black-box and white-box applicability~\cite{MIA1}. This white-box attack works by trying to predict the most likely input for a given label. This allows the adversary to reconstruct a generalized input image for a specific label.

The model inversion attack is only applicable in linear models. Furthermore, the authors note that it is often quite easy to come up with models for which this attack method does not work. A big weakness of this attack is that it is computationally infeasible for large input spaces since it in essence brute-forces all input combinations. 

\subsubsection{mGAN-AI}
The multitask GAN for Auxiliary Identification (mGAN-AI) attack relies heavily on the training of GANs to reconstruct accurate approximate training samples. This method of attack requires the model updates from each client and an auxiliary dataset. From these updates, a set of fake input images is generated which result in the same update. Then a discriminator network tries to figure out: 1.~if these images are fake or real, 2.~to which client these images belong, and 3.~to which category the images belong.

After some training, the mGAN-AI can provide a revealing reconstruction of training samples belonging to a victim client. In addition to this passive attack, the authors also describe an active variant which works in the same way but ensures that non-clients don not participate. This allows the GAN to optimize its reproduction capabilities towards a targeted victim~\cite{mGAN-AI}. Demonstrative results show that this technique is very viable to compromise client-level privacy.

This method of attack is only applicable in cases where clients have mostly homogeneous data and there is an auxiliary dataset available. Since the method relies on generative adversarial networks, this method is only applicable on data which can be synthesized.

\subsubsection{GAN}
This attack actively targets a victim client to release private data by poisoning the shared model in the adversaries favour. In a simple setting, the adversary uses the shared model to differentiate the victim's model update from which the adversary learns which labels have been used. Then the adversary uses a generative adversarial network (GAN)~\cite{GAN} to generate images for one of the labels exclusively used by the victim. The adversary then purposefully mislabels these images as one of the labels exclusively used by itself. The gradients of the victim will become steeper as the generated and subsequently mislabeled images get more re-presentable as the victim's images~\cite{GAN-attack}.

There is some critique to this attack. Although useful in a controlled environment, this attack method supposes that the victim and adversary share at least one shared label and one exclusive mutual label which is unrealistic as the client pool gets larger. Additionally, it is infeasible to distinguish a targeted victim in a practical federated setting due to the randomness of the selected clients and the degradation of influence of the adversary when averaging weights~\cite{mGAN-AI}.

\subsection{Attacks targeting misclassification}
Here we discuss only \emph{adversarial example attacks} as a form of misclassification attacks.
Adversarial example attacks aim to produce samples that will be purposefully misclassified at runtime \cite{privacy_ml, threat_of_ad_in_cv}. Federated learning is vulnerable to this because the deployment of the model means that adversaries have virtually no restrictions when crafting adversarial examples. If the intended use of the model is, for example, on-device authentication (e.g. by means of face recognition) then adversarial example attacks can be very effective.

\subsection{Attacks targeting model corruption}

Here we discuss only \emph{model poisoning attacks} as a form of model corruption attacks. Using this attack method, every client can directly influence the weights of the shared model which means that poisoning the shared model becomes trivial. Model poisoning can be used to introduce a backdoor which can be leveraged by the adversary~\cite{backdoor2,Backdoorattack,backdoor3,sybilattack}.

A specific attack method designed for FL was demonstrated by Bagdasaryan et al.~\cite{Backdoorattack}. The attack poisons the model by exploiting the fact that a client can change the entire shared model by declaring that it has a significantly large amount of samples. This allows a client to change the shared model in order to insert a backdoor. One example use case is forcing a shared model to be biased towards an adversarial label/prediction, such as forcing a word auto-completion model to suggest a brand name, or to force a model to incorrectly classify images of a political figure as inappropriate.

Similar work by Bhagoji et al.~\cite{modelpoison} demonstrate the viability of this method further and show how this attack method can be used in a stealth manner. Furthermore, they demonstrate how to circumvent byzantine-robust aggregation mechanisms~\cite{Byzantine-Robust1, Byzantine-Robust2}. These types of aggregation algorithms are designed specifically to reduce the influence of singular clients.

A different approach is taken by Fung et al.~\cite{sybilattack} where, rather than relying on boosting of sample numbers or gradients, the adversary is capable of joining and leaving different federated clients. This type of attack is called a sybil-based attack \cite{sybilattack,sybil}. The adversary can then return a multitude of poisoned models back to the server each iteration.

Closely related to sybil attacks are joint-effort model-poisoning attacks. Anecdotal evidence suggests a possible vulnerability for models that train on user input. For example, a twitter chat-bot developed by Microsoft that learned from user communication was corrupted by a coordinated effort of users to post racist and sexist tweets \cite{tay}.

In addition to model poisoning attacks, there are also data poisoning attacks. These types of attacks can be successfully used for adversarial goals such as flipping labels~\cite{labelflip, sybilattack}. However, it is noted that for FL all data poisoning attacks are inferior to model poisoning attacks for the simple reason that the data is not sent to the server. Therefore, anything that can be achieved with poisonous data is equally viable by poisoning the model~\cite{modelpoison, threatsurvey1}.

\section{Defensive measures} \label{sec:defense}

\begin{table*}[t]
\caption{This matrix illustrates the effectiveness of defensive measures against the attack surface as illustrated in Figure~\ref{fig:fed_steps}}
\label{tab:ASmatrix}
\centering
\begin{tabular}{@{} cl*{3}c|*{4}c | @{}}
&& \multicolumn{3}{c}{Server}&\multicolumn{4}{c}{Client} \\
&\multicolumn{1}{l|}{}&
\rot{\begin{tabular}{@{}l@{}}1) Manipulate\\ model per client\end{tabular}}&
\rot{\begin{tabular}{@{}l@{}}2) Access to client-\\specific model\end{tabular}}&
\rot{\begin{tabular}{@{}l@{}}3) Access to\\ aggregate model\end{tabular}}&
\rot{\begin{tabular}{@{}l@{}}1) Access to\\ aggregate model\end{tabular}}&

\rot{\begin{tabular}{@{}l@{}}2) Train on\\ adversarial data\end{tabular}}&
\rot{\begin{tabular}{@{}l@{}}3) Manipulate\\ gradient update\end{tabular}}&
\rot{\begin{tabular}{@{}l@{}}4) Influence\\ aggregation\end{tabular}}\\ \cline{2-9} 

\begin{tabular}{@{}l@{}}\end{tabular} 

\multirow{7}{*}{\rot{Defenses}}
&\multicolumn{1}{l|}{Gradient Subset}        &$-$           &$+$        &$\backsim$ &$-$        &$-$    &$-$        &$-$\\
&\multicolumn{1}{l|}{Gradient Compression}   &$\backsim$    &$\backsim$ &$-$        &$\backsim$ &$-$    &$-$        &$-$\\
&\multicolumn{1}{l|}{Dropout}                &$-$           &$\backsim$ &$-$        &$\backsim$ &$-$    &$\backsim$ &$\backsim$\\
&\multicolumn{1}{l|}{DP}                     &$-$           &$+$        &$\backsim$ &$\backsim$ &$-$    &$-$        &$-$\\ 
&\multicolumn{1}{l|}{SMC}                    &$-$           &$+$        &$-$        &$\backsim$ &$-$    &$-$        &$\backsim$\\
&\multicolumn{1}{l|}{Homomorphic encryption} &$+$           &$+$        &$+$        &$-$        &$-$    &$-$        &$-$\\
&\multicolumn{1}{l|}{Robust aggregation}     &$-$           &$-$        &$-$        &$-$        &$+$    &$+$        &$+$\\ 
\cline{2-9}

\end{tabular}
\end{table*}

\begin{table}[b]
\caption{A matrix that describes the effectiveness of defensive measures against attack methods. $+$: reliably effective, $\backsim$: limited effectiveness, $-$: the defensive measure may be presumed wholly ineffective or trivially circumvented, $\ast$: context dependent effectiveness.}
\label{tab:effectivenessmatrix}
\centering
\begin{tabular}{@{} cl*{8}c | @{}}
&& \multicolumn{8}{c}{Attacks} \\
&\multicolumn{1}{l|}{}&\rot{Loss-function/Relu}&\rot{First dense layer}&\rot{DLG/iDLG}&\rot{MIA}&\rot{mGAN-AI}&\rot{GAN}& \rot{Adversarial example} &\rot{Model poisoning}\\ \cline{2-10} 

\multirow{7}{*}{\rot{Defenses}}
&\multicolumn{1}{l|}{Gradient Subset}        &$+$        &$\backsim$ &$\backsim$ &$+$           &$\backsim$ &$\backsim$ &$-$        &$-$\\
&\multicolumn{1}{l|}{Gradient Comp}          &$\backsim$ &$-$        &$-$        &$\backsim$    &$\backsim$ &$\backsim$ &$\backsim$ &$-$\\
&\multicolumn{1}{l|}{Dropout}                &$+$        &$-$        &$-$        &$+$           &$\backsim$ &$+$        &$+$        &$\backsim$\\
&\multicolumn{1}{l|}{DP}                     &$+$        &$\ast$     &$+$        &$+$           &$\ast$     &$+$        &$+$        &$-$\\ 
&\multicolumn{1}{l|}{SMC}                    &$-$        &$\backsim$ &$\backsim$ &$\backsim$    &$+$        &$-$        &$-$        &$-$\\
&\multicolumn{1}{l|}{Homomorphic enc}        &$+$        &$+$        &$+$        &$+$           &$+$        &$-$        &$-$        &$-$\\
&\multicolumn{1}{l|}{Robust aggregation}     &$-$        &$-$        &$-$        &$\backsim$    &$-$        &$\backsim$ &$-$        &$+$\\
\cline{2-10}

\end{tabular}
\end{table}



This section describes various strategies proposed in the literature for increasing the security of the basic FL algorithm~\cite{FedLearning}. Defensive measures found in multiple works of literature are group by their underlying defensive strategy. Table~\ref{tab:effectivenessmatrix} lists the defensive measures that are discussed in this section and displays their effectiveness against the aforementioned attack methods. Different strategies will prove more valuable to thwart different attack methods and generally a combination of defensive measures is required for optimal defensiveness. In this table, the $\backsim$ symbol represents limited effectiveness of the defensive measure. This means that the attacker needs to perform extra steps to achieve its goal reliably or that the attacker may not presume that the outcome of the attack is reliable (e.g. very noisy reconstructions). Furthermore, the $\ast$ symbol represents the cases where the context of implementation is crucially important for the evaluation of effectiveness. For example, to assure privacy protection with differential privacy, the amplitude of noise that is to be added may result in a non-converging collaboratively trained model. Whereas lesser amplitudes prompt viability of an attack method. Table~\ref{tab:ASmatrix} illustrates the added protection for the specific vulnerable points illustrated in~\ref{fig:fed_steps}.

\subsection{Gradient subset}
Initially proposed for the use of Distributed distributed stochastic gradient descent (DSGD), only a subset of all gradients should be communicated to improve communication efficiency~\cite{DSSGD}. The authors propose a selection method for communicating only the most important gradients as well as adding a randomized weight selection variant. Such a randomized variant is also proposed briefly for a federated setting by~\cite{FL_strategies_comm_eff} for increasing communication efficiency. Yoon et al.~\cite{APC} proposed a different gradient subset approach and subsequently demonstrated that sending sparse and selective gradients can improve performance for a non-IID dataset.

As a defensive measure, such methods reduce the available information the server has from any client. The incompleteness of such information may cause problems for some attack methods to reliably infer information from the update. However, recent studies show that such methods do not prevent adversaries from inferring reliable information from victims~\cite{Exploiting_Feature_Leakage, homomorphic}.

\subsection{Gradient compression}
Although mainly devised for the use of communication efficiency, lossy compression techniques may also be implemented to facilitate some form of information security. Konecny et al.~\cite{FL_strategies_comm_eff} demonstrate how such a method reduces communication costs. The server only has victim specific incomplete information.

Other forms of compression are also possible. An autoencoder compressor is proposed to encrypt data transfer~\cite{NN_defence} in which the server trains an autoencoder on dummy-gradient updates, and subsequently releases the encoding part to each of the clients whilst keeping the decoder part secretive. Every gradient update from each client is now encoded before transfer to the server and decoded server side. This has a couple of advantages. First, the model is compressed by the encoder which will increase communication efficiency. Second, the encoding and decoding result in some information loss about the gradient update. How well this measure prevents the server from malicious reconstruction remains to be seen.

\subsection{Dropout} 
Another way to try to increase defensiveness is to employ a technique called dropout~\cite{dropout}. Although generally used to prevent over-fitting~\cite{dropout2}, dropout does introduce a certain (finite) randomness to the gradient updates. Training on a set of data-points will not have a deterministic gradient update associated with it, which reduces the exploitable attack surface. Due to the feature-generalizing nature of dropout, it could have an adverse effect if the goal is to infer generalized data rather than exact information. This was demonstrated by~\cite{Exploiting_Feature_Leakage}.


\subsection{Differential privacy (DP)}
Differential privacy~\cite{diffprivacy} aims to ensure privacy mathematically by introducing noise to variables in the system. DP is often regarded as one of the strongest privacy standards since it provides rigorous provability of privacy. DP has been successfully implemented on machine learning applications such as linear regression~\cite{regression_diff}, SVM's~\cite{SVM_diff}, and deep learning~\cite{dssg_diff, deep_withdiff}. One method to achieve this is to perturb the model before the optimization step in the learning algorithm~\cite{diffprivriskmin}. Others apply the perturbation at different stages of the SGD algorithm~\cite{diffprivSGD,deep_withdiff}.
However, it is challenging to implement DP in an FL context. DP works best when each client has access to a significantly large dataset~\cite{diffprivSGD}. Since the available data can vary in size between clients, the DP sensitivity should be considered on a per-client basis. Bonawitz et al.~\cite{diff_lang} proposed a DP implementation for federated averaging for an LSTM language model by clipping gradient updates and adding the noise at the server-side. This ensures that clients cannot infer or attack other clients but may not guarantee safety against server-side attacks. Geyer et al.~\cite{fed_diff} expanded on this work by implementing FL with DP on an image recognition network. Triastcyn et al.~\cite{Fed_dif_bays} demonstrated FL with Bayesian DP which may provide faster convergence in setting where the data is similarly distributed over participating clients. Bayesian DP leverages this distribution to justify lesser noise amplitude while still retaining the same or even tighter privacy bounds.

\subsection{Secure multiparty computation (SMC)}
Secure multiparty computation schemes are algorithms that allow two or more participants to jointly compute functions over their collective data without disclosing any data to the other party. Using such methods to securely aggregate the gradient models from individual clients can be used to shield clients from the server.

Common SMC schemes are developed for the use in secure two-party computation in which there are only two communicating/computing entities. Such schemes have also been developed for the use of joined machine learning~\cite{SMC_2pc,SMC_SecureML,SMC_fair,privpy}. In terms of federated learning, this often requires two untrusting servers that connect to some or all the participating clients.

Other than two-party computation, three-party communication have been proposed as-well~\cite{Aby3,3PC_1,3PC_2,3PC_3}. These methods rely on at least semi-honest participants or an honest majority. Mohassel et al.~\cite{Aby3} proposed a novel solution in which three servers use SMC to compute ML-model aggregates. They additionally hint towards a structure in which the clients are subdivided and represented by different servers. If the server is to be trusted, such a structure can even be extended to a hierarchical structure in which a server represents several underlying servers.

For federated learning, one key weakness is that the server can know exactly a specific clients gradient update and thereby might infer information about a targeted victim client. Bonawitz et al.~\cite{prac_sec_agg} proposed a combination of several protocols to ensure security in a practical federated setting through obscuring the aggregation from the server.


\subsection{Homomorphic encryption}
Homomorphic encryption relies on a mathematical encryption method in which mathematical operations applied on an encrypted message result in the same mathematical operation being applied to the original message when the message is decrypted. Using homomorphic encryption in federated learning theoretically ensures no performance loss in terms of model convergence~\cite{homomorphic}. 

Homomorphic encryption comes in three forms: \emph{Partially Homomorphic Encryption (PHE)} which allow only a single type of operation. \emph{Somewhat Homomorphic Encryption (SWHE)} which allow multiple mathematical operations for only a bounded number of applications. \emph{Fully Homomorphic Encryption (FHE)} which allows unlimited amounts of operations without restriction \cite{Homomorphic_survey}.
Homomorphic encryption is especially suitable for ensuring privacy against an untrustworthy server. Such a method was proposed by~\cite{First-layer-attack} to secure DSGD but may easily be adapted to work on federated averaging. Homomorphic encryption has been proposed as a means to ensure privacy for medical data as-well~\cite{homomorphic_medical,  homomorphic_medical2, Homomorph}. Dowlin et al.~\cite{cryptonets} proposed an implementation for learning on encrypted data with so-called \emph{cryptonets}. Phong et al.\ demonstrated a simple yet effective additive homomorphic scheme for federated learning~\cite{homomorphic}.

Other collaborative learning strategies include the use of homomorphic encryption to encrypt data and allow a centralized server to train on this encrypted data~\cite{homomorphic_deep,homomorphic_backprop}

Homomorphic encryption does have some major drawbacks that are especially relevant for large scale implementation. First, practically viable fully homomorphic encryption such as FHE generally has large computation overhead which often makes its implementation impractical~\cite{privpy, Homomorph, Homomorphic_survey}. Second, for SWHE the data-size of the encrypted models increase linearly with each homomorphic operation~\cite{Homomorphic_survey}. Thus the encrypted models are notably larger than the plain model, thereby increasing communication costs. Third, homomorphic encryption requires communication between participating clients to facilitate key-sharing protocols. Communication between clients is not desired in FL and this also introduces a possible vulnerability if new clients should be able to join in-between training rounds. Fourth, without additional safeguards, homomorphic encryption is still vulnerable to all client-side attacks. Additionally, problems may arise for the evaluation of the model. Lastly, it should be noted that the server in a proper homomorphic scheme does not have direct access to the jointly trained model, which is (generally) the reason for the server to do FL in the first place. 

\subsection{Robust aggregation}
In the regular FL setting~\cite{FedLearning} a client sends the newly trained model as well as the number of samples it has back to the server. Since the client can alter these values more robust aggregation methods are required that prohibit malicious clients from exploiting the weighted averaging approach (in which the weights are determined by clients local data size). Furthermore, multiple clients may work together to conceal their malicious efforts. The centralized server has no way of knowing which of the clients are malicious. This is often revered to as problems involving Byzantine clients~\cite{byzantine}.

Blanchard et al.~\cite{Byzantine-Robust1} addressed this problem by proposing an aggregation rule called~\emph{Krum} which minimizes the sum of squared distances over the model updates. Effectively removing outliers that are supposedly provided by byzantine clients. Hereby the aggregation is secure against at most less than $\dfrac{n-2}{2}$ byzantine workers among $n$ clients. Although this method is exploitable still ~\cite{bulyan}.

El Mhamdi et al.~\cite{bulyan} proposed a byzantine–resilient aggregation rule which is to be used in combination with other rules named~\emph{Bulyan} for the use in DSGD. Hereby the aggregation is secure against at most $\dfrac{n-3}{4}$ byzantine workers among $n$ clients.

Li et al.~\cite{Aggregate_trust_weight} proposed a weighting strategy by reliability for the use of Heterogeneous Data. The 'source reliability' is determined by calculating the distance between the clients' update and the aggregate update. 

Pilluta et al.~\cite{fed_sec_aggr} proposed an approximate geometric median for the use in federated learning and demonstrated its effectiveness by introducing model corrupting updates and data. The results demonstrate only a substantial robustness improvement for linear models.

Harry Hsu et al.~\cite{FedAvgM} proposed a server-side momentum implementation specifically for the use of non-IID client datasets. Combined with the random client selection native to federated learning~\cite{FedLearning} this may greatly reduce the influence a malicious client has over a multitude of rounds.

Chang et al.~\cite{Cronus} devised a clever knowledge-sharing algorithm for robust privacy-preserving, and heterogeneous federated learning named~\emph{Cronus}. Their algorithm relies on a publicly available dataset known to all clients. Instead of sending model updates, participating clients send their predicted labels from the publicly available samples. Their subsequent aggregation uses the~\emph{robust mean estimation}~\cite{Robust_mean} algorithm which is specifically developed or high dimensional robust aggregation. Hereafter the aggregated labels are sent to all clients, which subsequently use these labels to update their local model. Cronus has been demonstrated to be consistently robust against basic model poisoning strategies at the cost of model accuracy.


\section{Related work} \label{sec:related}
There is a small number of papers available that discuss different attack methods and defensive strategies for FL models. A recent survey paper~\cite{threatsurvey1} only lists attack methods and assesses their value on the implementation method of the federated algorithm. Other works on federated learning such as~\cite{related1, related2} list various defensive strategies but are more focused on implementation and application consideration of FL. Xu et al.~\cite{HybridAlpha} provides a taxonomy by which to list defensive strategies and subsequently shows a small number of defensive strategies categorized. Kairouz et al.~\cite{related3} provide an elaborate description of open problems in FL. These problems are accompanied by a detailed description of privacy concerns in terms of exploitable vulnerabilities and defenses.


\section{Conclusion} \label{sec:conc}
This paper discussed the basic principles of FL and showed multiple vulnerabilities FL has to insider attacks. A client cannot assume that data privacy is preserved since the centralized server can exploit these vulnerabilities to obtain information about private client data. At the same time, several security-enhancing strategies can be implemented. Some of which provide a mathematical proof of privacy preservation whilst others show their effectiveness towards a specific attacking strategy. The paper further lists the attack methods found in the literature. These attacks are classified using several characteristics: the source of attack, active and passive attacks, white-box or black-box viability, the attacker's goal, and further restrictions.
Furthermore, defensive measures are listed by their underlying principle. An easy to use matrix is illustrated that provides a rudimentary evaluation of the effectiveness of defensive methods against attack strategies.

\balance
\bibliographystyle{unsrt}
\bibliography{References}
\end{document}